\documentclass[12pt]{iopart}
\usepackage{graphicx}
\usepackage{iopams}  

\usepackage{color}
\usepackage{xcolor}
\usepackage{ulem} 
\usepackage{upgreek}

\begin{document}

\title[Magnetic field characterization of edge currents in quantum spin Hall insulators]{Magnetic field Characterization of edge currents in quantum spin Hall insulators}

\author{Felipe Pinto$^{1}$, Ricardo C Heitzer $^{1}$, Eitan Dvorquez $^{1}$, Roberto Rodriguez $^{1}$,  Qiang Sun $^{2}$, Andrew D Greentree $^{2}$, Brant C Gibson $^{2}$ and Jerónimo R Maze$^{1}$}

\address{$^1$Department of Physics, Pontificia Universidad Católica de Chile, Vicuña Mackenna Avenue 4860, 8940000, Santiago, Chile.}
\address{$^2$ARC Centre of Excellence for Nanoscale Biophotonics, RMIT University, Melbourne, VIC 3001, Australia.}

\vspace{10pt}
\begin{indented}
\item[]September 2024 
\end{indented}

\begin{abstract}

Quantum spin Hall (QSH) insulators are materials with nontrivial topological properties, characterized by helical edge currents. In 2D strips, the application of a bias voltage along the edge generates a magnetization that can be measured using quantum sensors and magnetometry techniques. In this work, we calculate the magnetic field in the vicinity of the edge and explore the potential role of nitrogen-vacancy (NV) centers in diamond as local probes for the characterization of QSH edge states in topological insulators. We characterize the magnetic field near the edges produced by both electron currents and spin accumulation at the edge. We focus on identifying the position from the edge at which the effects of spin accumulation become detectable. We observe that a larger gap between the conduction and valence bands, along with a lower Fermi velocity, results in a stronger magnetic field, with the detectable spin accumulation being more concentrated near the edge.  Conversely, a smaller gap results in a slight reduction in the magnetic field magnitude, but the field associated with spin accumulation becomes detectable further from the edge. This work provides insights that could be useful for the characterization of topological materials and the development of novel electro-optical devices.

\end{abstract}

%
\vspace{2pc}
\noindent{\it Keywords}: Topological insulators, QSH effect, NV magnetometry, Quantum sensing.
%
%
%
%
\section{Introduction}

Topological insulators have gained great relevance in the last decades due to their  applications in advanced nano-electronic and spintronic devices \citation{app1,app2,app3,app4}. This new class of solid-state materials exhibits a bulk band gap like an ordinary insulator but  protected conducting states on their edge or surface, arising from the topology of the material. 

Quantum spin Hall (QSH) insulators are related to topological insulators as they are also characterized by a topological invariant. The QSH case exhibits a time-reversal invariant  Hamiltonian, presenting edge states with opposite spin polarization propagation \citation{TI1,TI2}. Currently, new materials have been reported as QSH insulators that exhibit gaps of the order of 0.1 eV such as graphene-based materials incorporating transition metals like ZrC monolayers and OsC monolayers  \citation{Reals_TI,Reals_TI1,Reals_TI2}, among others \citation{Reals_TI}. These properties make them very attractive for fabricating high-mobility and non-dissipative electrical devices \citation{app1,TI1,TI2,TI3,TI4}.

With the development of nanoscale magnetometry using atomic-sized sensors such as nitrogen-vacancy (NV) centers in nanodiamonds, we can extract information within a few nanometers of the system under study. These sensors benefit from a large toolbox of spin manipulation techniques to perform nanoscale magnetometry \citation{NV1, NV2, NV3, NV4, NV5, NV6}. NV centers have been widely used in detecting magnetic textures from both electric currents and spins \citation{NV4, NV_sensing1, NV_sensing2}. These centers are particularly suitable for this task due to their high coherence times, even at room temperature \citation{NV1, NV2, NV3}. Consequently, the use of these sensors on the tips of atomic force microscopes allows for the scanning of magnetic fields of various samples. In particular, their sensitivity can characterize the weak magnetic fields present at the edges of topological insulators, enabling their local study with minimal intervention.
\begin{figure}[tb!]
\centering
\noindent{\includegraphics[width=.6\textwidth]{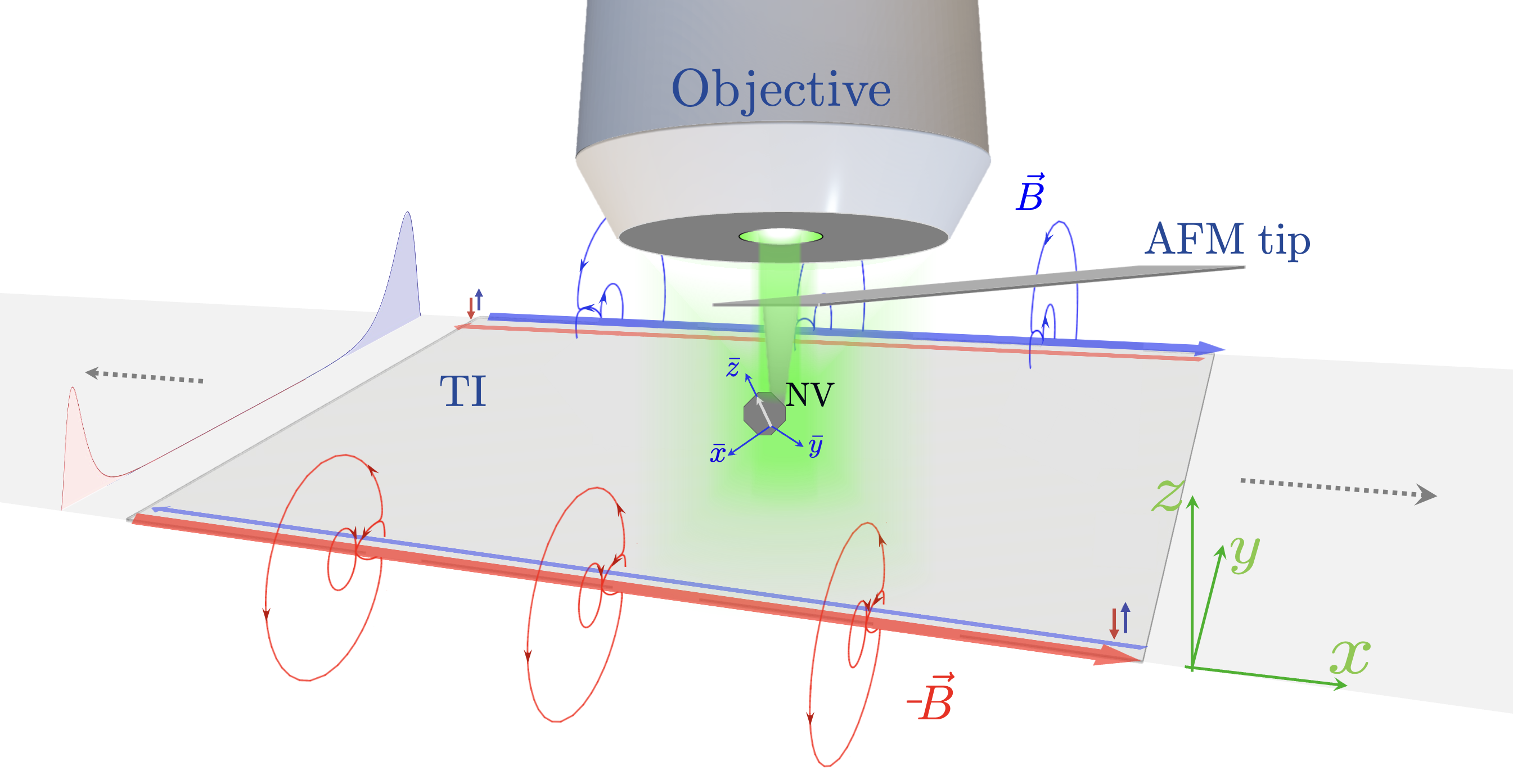} 
\caption[]{Schematic representation of a topological insulator and an NV center based sensor on a scanning AFM probe. Conductive edge states are depicted in colour along the top and bottom sides of the TI, along with their probability density across the $y$-axis, highlighting that for a given spin, one edge state becomes more populated when an electric potential is applied. The magnetic field near the edge is illustrated by field lines. An objective lens  records the NV fluorescence.}\label{SetUpPicture}}
\end{figure}

In this work, we calculate the magnetic field generated near the edge of a 2D topological insulator exhibiting the quantum spin Hall (QSH) effect and demonstrate the feasibility of identifying the QSH phase by performing calculations sufficiently close to the sample under the application of a weak electric field. In this phase, spin-locking leads to spin accumulation at the edge \citation{TI1,TI2,TI4}, which modifies the magnetic field generated by the electron current. Using linear response theory, we calculate the magnetic field produced by this spin accumulation at the edge, finding that it varies as \(1/d^2\), where \(d\) is the distance from the edge. This contrasts with the magnetic field generated by the current, which varies as \(1/d\). Therefore, in the region where both fields are comparable, a detectable change can be observed in both  magnitude and direction of the magnetic field. Finally, we emphasize that if this region lies within the operational range of NV centers in nanodiamonds, it is possible to measure this change in the magnetic field using magnetometry techniques.

This manuscript is organized as follows. Section~\ref{sect:one} describes the magnetic field near the edge due to electron current and spin accumulation. Section~\ref{sec:three} examines the dependence of the magnetic field on topological insulator parameters, such as the gap width and the Fermi velocity of the edge electrons, at distances where both contributions to the magnetic field are comparable. Section~\ref{sec:two} presents methods for characterizing this field using NV center magnetometry techniques. It also details the required sensitivity and probe properties needed to accurately characterize the magnetic field discussed herein. Finally, Section~\ref{sec:five} provides a discussion on the feasibility of measuring this quantum spin Hall effect in real topological insulators and offers considerations for future improvements.

\section{Edge Magnetic Field from Spin Accumulation and Electron Current }
\label{sect:one}

The simplest Hamiltonian representing a QSH insulator can be written in continuous form as \citation{TI5,Model1,Model2}

\begin{equation}
H(k_x,k_y)=\pmatrix{
A(k) \bi{I}_2+\bi{d}(k)\cdot\bi{\sigma} & 0 \cr
0 & A(k) \bi{I}_2+\left(\bi{d}(-k)\cdot\bi{\sigma}\right)^{\dagger}}\;,
\end{equation}\label{TI_Hamiltonian}

\noindent where \(\bi{I}_2\) is a \(2 \times 2\) unit matrix,  \(\bi{\sigma}\) are the Pauli matrices, \(A(k) = -A(k_x^2 + k_y^2)\), with \(A\) representing the asymmetry between the valence and conduction bands, \(d_x(k_x) = v_F \hbar \:k_x\) and \(d_y(k_y) = v_F \hbar \:k_y\), with \(v_F\) representing the Fermi velocity, and \(d_z(k_x, k_y) = M - C(k_x^2 + k_y^2)\), where \(2M\) denotes the band gap at \(k = 0\) and \(C\) is a parameter related to the band curvature. This model is highly versatile and sufficient for our purposes, as it allows us to modify both the band gap  and the Fermi velocity.

\begin{figure}[t]
\centering
\noindent{\includegraphics[width=.9
\textwidth]{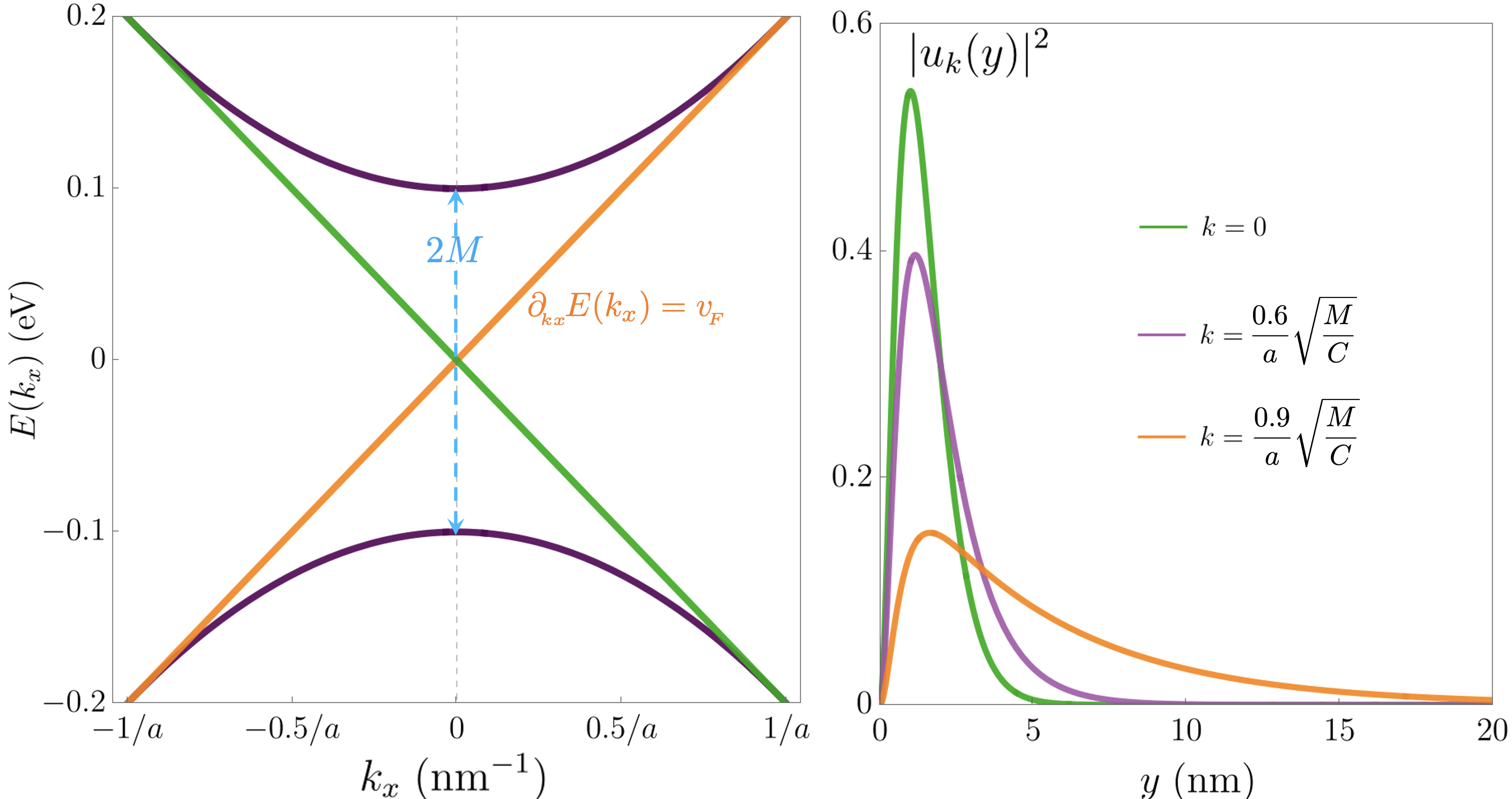}} 
\caption[]{(a) Dispersion relation for \(A=0\), \(M=-0.1\) eV, \(v_F\hbar= 0.5\) nm$^{-1}$ eV, and \(C=-0.5\) nm$^{-2}$ eV, highlighting the band gap width as \(2M\) and the Fermi velocity as the slope of the dispersion curves. (b) Cross-sectional profiles of the edge states penetrating into the bulk for three values of \(k\) within the range \(-\sqrt{\frac{M}{C}} < k < \sqrt{\frac{M}{C}}\), showing the spatial region where the probability density of the edge state is non-zero in this model.}\label{Dispersion relation}
\end{figure}

If we consider the material to be finite along the \(y\)-axis and periodic along the \(x\)-axis, this Hamiltonian exhibits conductive edge states when the condition sign\((MC)=1\) is satisfied. The dispersion relation for the conduction and valence bands, depicted in Fig.$\:$\ref{Dispersion relation} for given parameters \(A\), \(M\), \(v_F\), and \(C\), clearly shows conductive edge states, which are counter-propagating and helical on each edge of the strip, with the analytical form given by

\begin{equation}
    \psi_{k}^{\uparrow}(x,y) = \frac{1}{\sqrt{L}} e^{ikx} u_k(y) |\uparrow\rangle\;,
\end{equation}
\noindent where
\begin{equation*}
    u_k(y) = \frac{1}{\sqrt{N}} \left( e^{-\lambda_1(k) y} - e^{-\lambda_2(k) y} \right)\;,
\end{equation*}
\noindent and

\begin{equation*}
    N = \frac{\left( \lambda_1(k) - \lambda_2(k) \right)^2}{2 \lambda_1(k) \lambda_2(k) \left( \lambda_1(k) + \lambda_2(k) \right)}\;,
\end{equation*}

\noindent and \(L\) is the length of the strip. It is important to note that we are considering the strip to be sufficiently wide to make the interaction between edge states negligible \citation{TI6}, thus focusing on a single edge. Since the edge states on the same edge form Kramers pairs, the states can be related by \(\psi_{k}^{\downarrow}(x,y) = \Theta \psi_{k}^{\uparrow}(x,y)\), where \(\Theta = i\sigma_y K\) is the time-reversal operator consisting of a Pauli matrix \(\sigma_y\) and a complex conjugation operator \(K\).

\begin{figure}[t]
\centering
\noindent{\includegraphics[width=.6\textwidth]{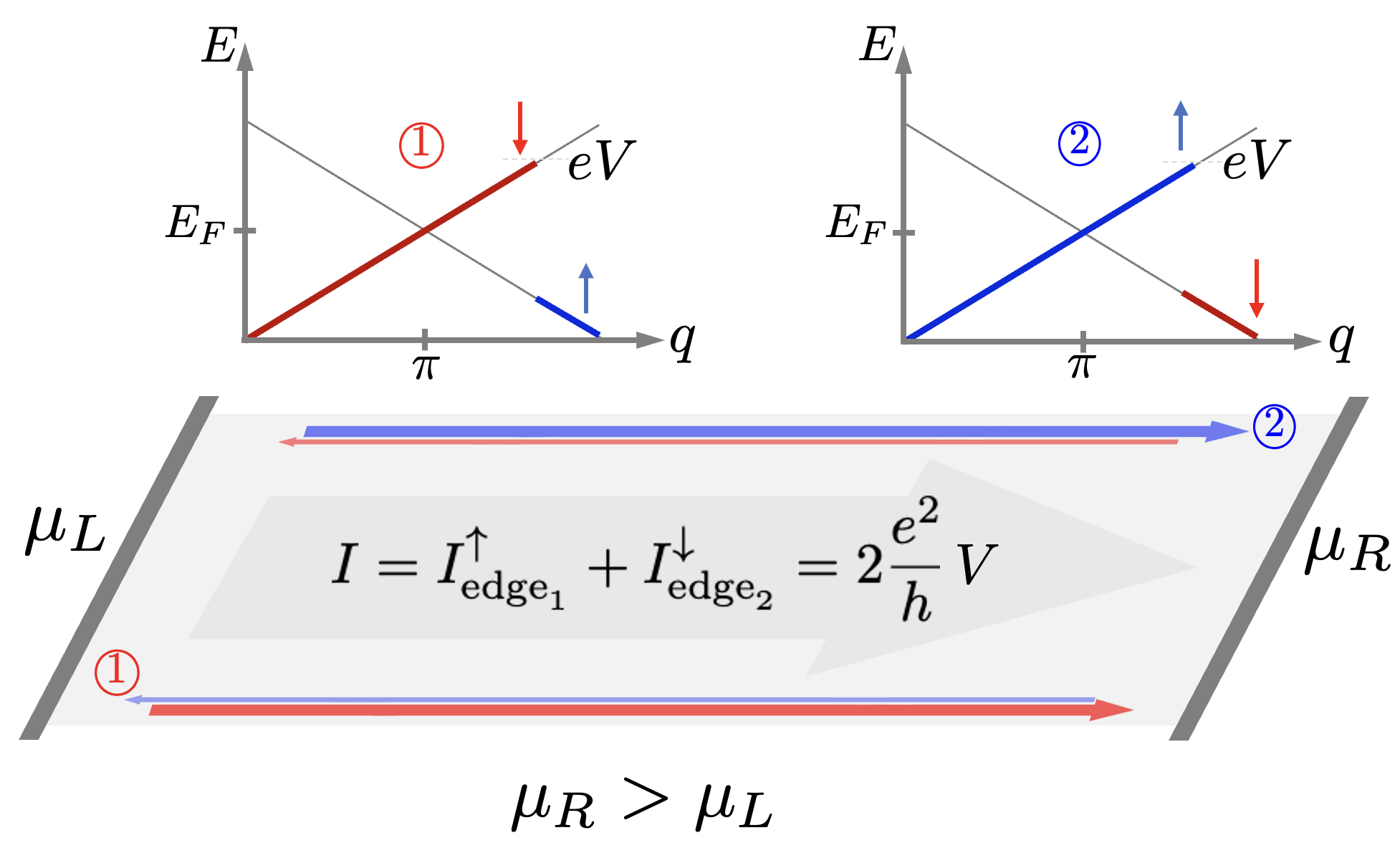}} 
\caption[]{At low temperatures, a pair of helical edge states with unequal populations is observed when connected to leads with chemical potentials $\mu_L$ and $\mu_R$. In this regime, the ballistic transport through the edge states is described by the Landauer-Büttiker framework \citation{TI_dispersion1,TIBook1}}\label{Current}
\end{figure}

For a two-terminal device of a topological insulator in the quantum spin Hall phase, due to the spin-locking of edge states, when a small electric field is applied along the edge as a perturbation to the system, an inverse magnetization at each edge is generated, as shown in Fig. \ref{Current}, although the sample as a whole is not magnetized \citation{TI1,TI2,TIBook1,TIbook2}. Thus, using linear response theory, we calculated the magnetization induced by the electric potential at a specific edge. The magnitude of this magnetization is given by (see \ref{sec:A})

\begin{equation}\label{main_magnetization}
m(q,y)=\frac{\mu_B}{\pi }\;\frac{ Ve}{k_B T} \;\delta(q)\hspace{-.3cm}\int\limits_{-\pi/a}^{\pi/a}\hspace{-.2cm}dk\;|u_k(y)|^{2}\frac{e^{v_{F}\hbar\: k\:\beta}}{\left(1+e^{v_{F}\hbar \beta\: k}\right)^2}\;,
\end{equation}

\noindent where \(\beta=1/k_BT\), \(\mu_B\) is the Bohr magneton, \(e\) is the electron charge and \(V\) is the applied voltage. In the $y$-direction, the magnitude of the magnetization is weighted by the edge state's probability density  $|u_k(y)|^{2}$,  and the delta function \(\delta(q)\) arises from the Fourier transform along $x$-axis of the spatial dependence of the voltage in the limit \(L \rightarrow \infty\). In real QSH insulators, the spin axis direction and hence the magnetization are strongly influenced by spin-orbit interactions and the specific material properties \citation{TI4,TI_WTe}. Generally, there is a competition between intrinsic and Rashba spin-orbit interactions \citation{TI3}. In this work, we assume that inversion symmetry is preserved in the material, neglecting the Rashba spin-orbit interaction . Consequently, the magnetization direction is expected to be in the plane perpendicular to the sample plane \citation{TI3,TI4,TI_HgTe}, although the orientation within this perpendicular plane may vary depending on the specific material \citation{TI_WTe}. With this magnetization, it is straightforward to obtain the magnetic field due to spin accumulation as

\begin{equation}
B_{\textrm{{\tiny spin}}}^{i}(\bi{r}) = \frac{\mu_0}{4\pi a^{2}} \hspace{-.1cm}\int\limits_{\hspace{-.2cm}-\pi/a}^{\hspace{-.1cm}\pi/a} \hspace{-.2cm}\int\limits_{0}^{\hspace{.2cm}L_y} \hspace{-.2cm}dq \, dy' \,e^{iqx} \, D_{iz}(q, y - y', z)  \;m(q, y')\;,
\end{equation}

\noindent where \(a\) is the lattice constant and \(D_{iz}(q, y - y', z)\) is a dimensionless parameter obtained as the semi-Fourier transform along the \(x\)-axis of the tensor \(D_{ij}(\bi{r} - \bi{r}') = \partial_{i} \partial_{j} \frac{a^{2}}{|\bi{r} - \bi{r}'|}\), where $\bi{r}=(x,y,z)$ and $i,j=x,y,z$. The magnetic field associated with the current can be directly calculated using the Biot-Savart law

\begin{equation}
\bi{B}_{\textrm{{\tiny current}}}(\bi{r}) = \frac{\mu_0 I}{2\pi} \int\limits_{\hspace{-.2cm}-\pi/a}^{\hspace{-.1cm}\pi/a} \hspace{-.2cm}\int\limits_{0}^{\hspace{.2cm}L_y} \hspace{-.2cm} dq \, dy' \, \frac{|u_q(y')|^{2}}{(y - y')^{2} + z^{2}}\pmatrix{0\cr z \cr (y'-y)}\;,
\end{equation}
\noindent where \(I = \frac{e^{2}}{h}V\) is the current generated in the low-temperature limit when the conductance is \(e^{2}/h\) \citation{TI4}. The temperature dependence of the current is complex and beyond the scope of this work. However, some studies suggest that as long as the thermal excitation energy is much lower than the energy gap between the conduction and valence bands, this assumption remains fairly accurate \citation{TIfiniteTemp, TIfiniteTemp2}.

\begin{figure}[t]
    \centering
    \noindent{\includegraphics[width=1\textwidth]{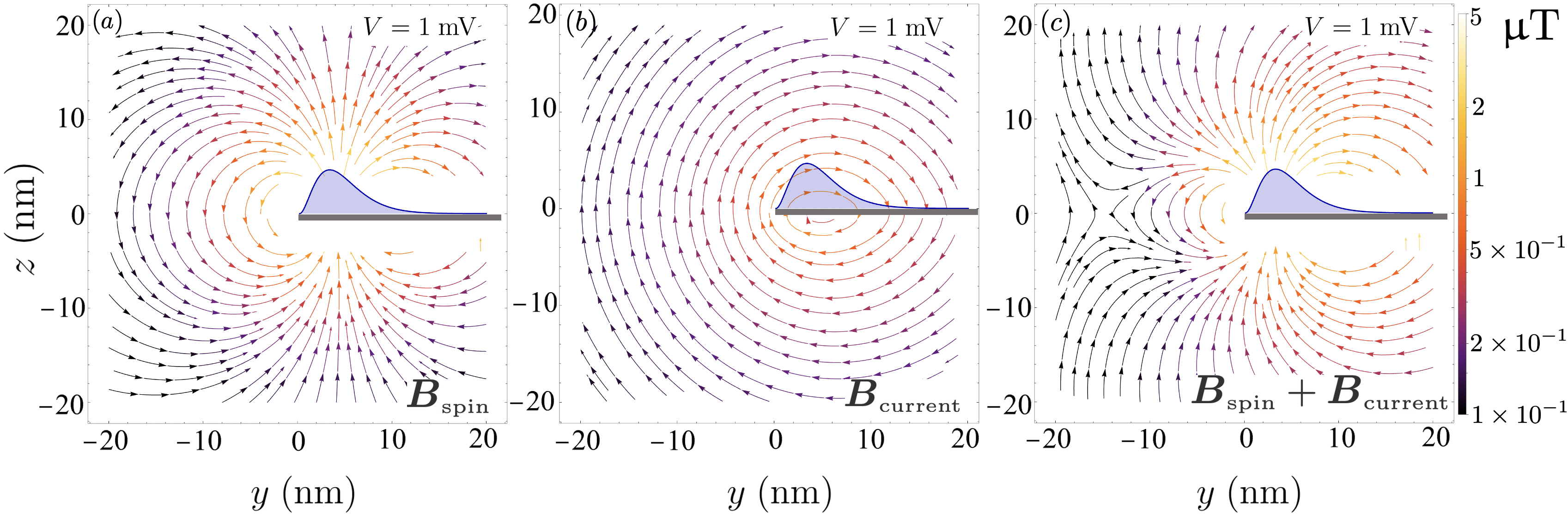}} 
    \caption[]{Cross-sectional magnetic field lines generated in a QSH insulator strip, with different colors indicating the field intensity. Panel (a) depicts the magnetic field lines arising from spin accumulation at the edge, panel (b) shows the magnetic field lines generated by the current, and panel (c) presents the superposition of both fields. The employed parameters are \(A=0\), \(\textrm{Gap}=0.1\) eV, \(v_F\hbar/a= 0.3\) eV, and \(C=-0.5\) nm\(^{-2}\) eV.}\label{FieldLines}
\end{figure}

 Figure \ref{FieldLines} shows the magnetic field generated by the spin accumulation, the current, and the sum of these two contributions for a particular topological insulator subjected to an applied voltage of 1 mV. Clear differences in the magnetic field profiles of both contributions can be observed. Above the edge of the topological insulator, the magnetic field \(B_{\mathrm{spin}}\) exhibits a significant component perpendicular to the material plane, while \(B_{\mathrm{current}}\) primarily points parallel to the plane. In the plane of the material but outside of it, \(B_{\mathrm{spin}}\) and \(B_{\mathrm{current}}\) have approximately opposite senses. Notably, there exists a point in this region where both contributions cancel each other out, as illustrated in Fig. \ref{FieldLines}c. Additionally, the decay behavior of each contribution varies with increasing distance from the topological insulator edge: \(B_{\mathrm{spin}}\) decays as \(1/d^2\), whereas \(B_{\mathrm{current}}\) decays as \(1/d\). In the following sections we further analyze these differences as a way to characterize the spin accumulation of topological materials.

\section{Magnetic Field Dependence on material properties}\label{sec:three}

\begin{figure}[t]
\centering
\noindent{\includegraphics[width=1\textwidth]{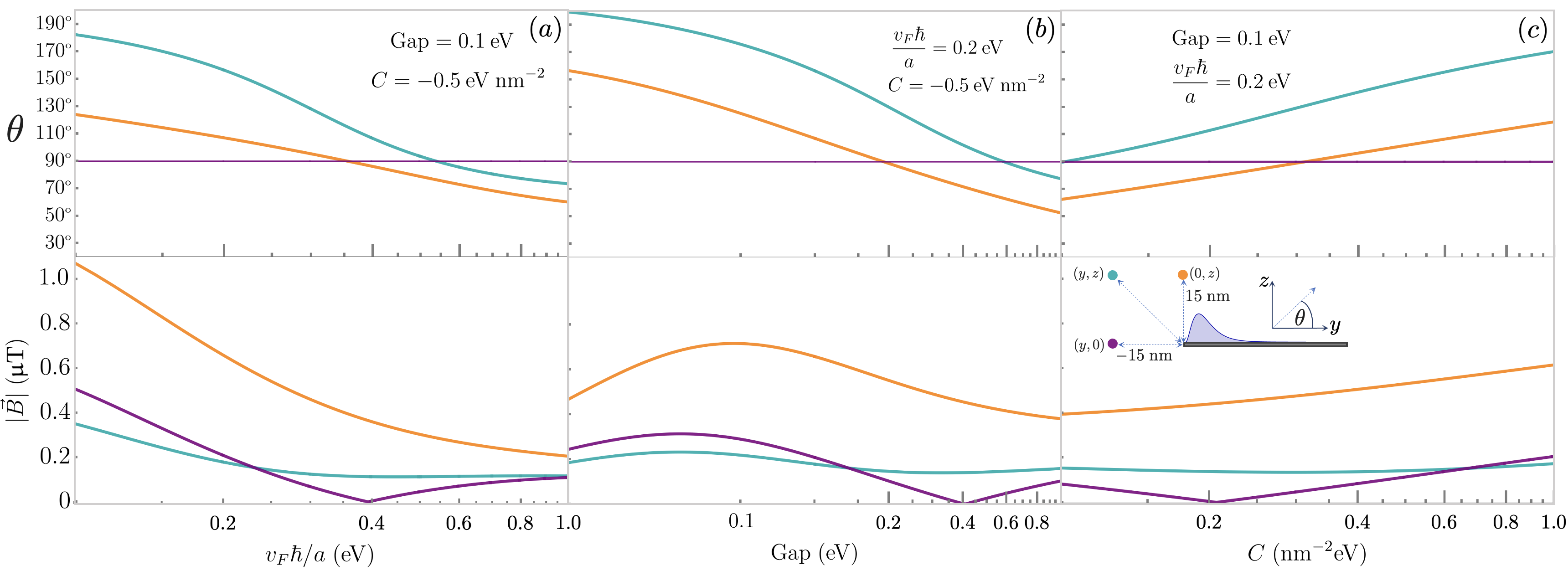}} 
\caption[]{Magnitude and and angle with respect to the $y$ axis of the total magnetic field as a function of the topological insulator parameters for three spatial points shown in the inset. Calculations were performed at a voltage \(V = 1\) mV and temperature \(T = 50\) K. Panel (a) displays the dependence on Fermi velocity, panel (b) shows the dependence on the band gap, and panel (c) illustrates the effect of the parameter \(C\). The lattice constant used in these simulations was \(a = 3.5\) Å.}\label{BvsvM}
\end{figure}

In the quantum spin Hall (QSH) phase, contributions to the magnetic field arise from both the electron current and spin accumulation at the edge. The latter being a result of spin-locking of the conducting states. This behavior is sensitive to the topological insulator's parameters, temperature, and voltage. The temperature effects are illustrated in Figure \ref{BvsT}, showing that at low temperatures, the magnetic field remains nearly constant, while at higher temperatures, the magnetic field effects due to spin accumulation diminish. Evidently, any effect from spin accumulation is erased when thermal energies approach the gap between the conduction and valence bands. Conversely, the voltage dependence is linear, as we are considering an electric potential energy two orders of magnitude lower than the gap between the conduction and valence bands of the topological insulator.

In relation to the magnetic field's dependence on the insulator parameters, Figure \ref{BvsvM} shows how the magnetic field’s magnitude shifts as the Fermi velocity, band gap, and band curvature \(C\) are varied. In panel (a), it is observed that as the Fermi velocity increases, the magnetic field magnitude decreases. This can be primarily attributed to the fact that, for a given voltage, a smaller number of states are populated at higher Fermi velocities, directly influencing the slope of the dispersion relation. A similar argument explains why higher Fermi velocities lessen the temperature's effect, as also shown in Figure \ref{BvsT}. In the same panel, the magnetic field angle decreases because the contribution of spin accumulation to the magnetic field is reduced.

In panel (b), an increase in the band gap causes the angle relative to the horizontal to change, even though the magnetic field magnitude remains largely unaffected. This is due to the fact that a larger gap confines the edge state closer to the boundary. Finally, in panel (c), the field’s magnitude and angle relative to the horizontal increase, as the parameter \(C\) affects the edge state in a manner opposite to that of the gap parameter.

\begin{figure}[t]
    \centering
    \noindent{\includegraphics[width=1\textwidth]{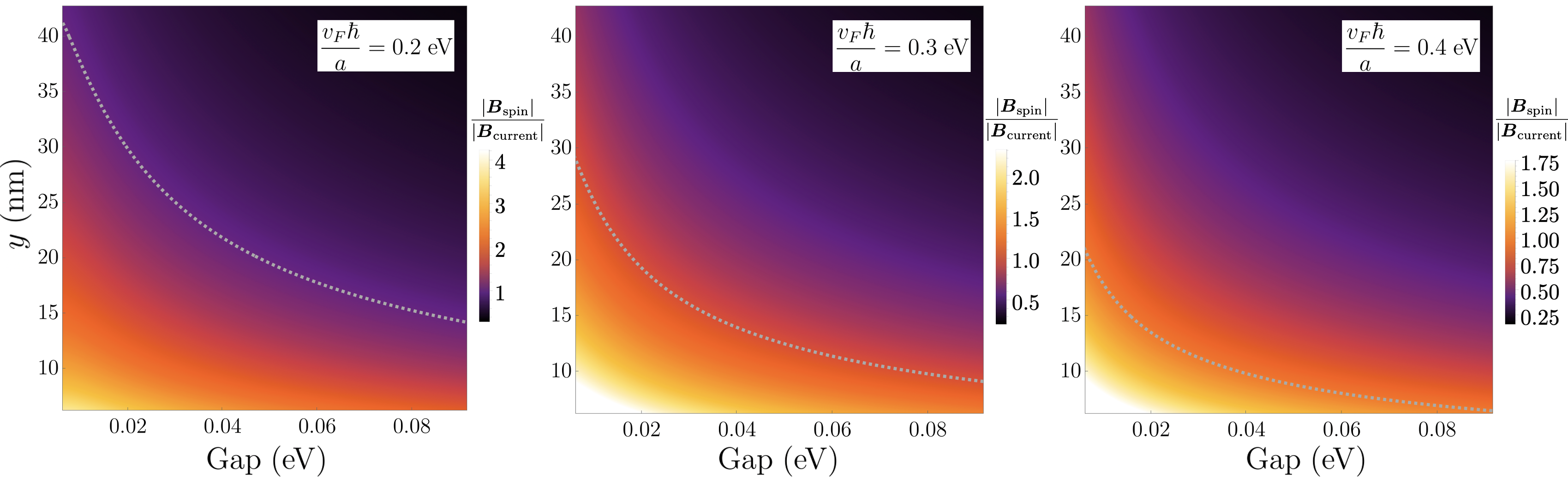}} 
    \caption[]{Ratio between the magnetic field magnitudes due to spin accumulation at the edge and  electron current as a function of the topological insulator gap and distance along the \(y\)-axis for \(z=0\) for three different Fermi's velocities. The dashed lines represent points at which both magnitudes are equal. The voltage used for the calculation of the magnetic fields was \(V= 1 \textrm{mV}\). The fixed parameter for the topological insulator was \(C=-0.5 \:\textrm{nm}^{-2}\textrm{eV}\), and the lattice constant was \(a=3.5\:\)\r{A}.}
    \label{Mvsy}
\end{figure}

Throughout the graphs in Figure \ref{BvsvM}, points of zero magnitude can be observed along the \(y\)-axis (indicated by the purple line), representing locations where the fields from the current and spin accumulation cancel each other out. Figure \ref{Mvsy} further illustrates the ratio between both contributions to the total magnetic field for an electric potential of 1 mV. This figure shows that as the gap increases, the magnetic field cancellation point moves closer to the edge, largely because the edge state is more localized near the boundary. This effect becomes even more pronounced as the Fermi velocity of the edge electrons increases.

\section{Magnetometry of Edge Spin Accomulation }\label{sec:two}

To characterize the magnetic fields generated by the edge state currents, NV centers in nanodiamonds or nanopillars can be employed. These sensors must be positioned close enough to the edge to differentiate the effects of spin accumulation. NV nanodiamonds with diameters smaller than 10 nm and stable photoluminescence, as well as average coherence times of \(T_2 = 1 \, \upmu\)s, have been reported in \citation{NV_sensing3,NV_sensing4,NV_sensing5}. The magnetic sensitivity of these sensors, of the order of 0.1 \(\upmu\)T Hz\(^{-1/2}\), would enable the detection of the magnetic fields discussed.

The proximity at which the effect due to spin accumulation becomes detectable depends on the parameters associated with the topological insulator, as illustrated in Figures \ref{BvsvM} to \ref{Mvsy}. The points where the magnetic fields from the current and spin accumulation are of the same order set the distance at which spin accumulation effects become detectable, thus establishing certain limits on the use of NV sensors in nanodiamonds. Furthermore, the mobility of the AFM microscope tips, which house the NV defects in nanodiamonds or nanopillars, allows detection in various regions around the edge, facilitating the potential reconstruction of the vector field associated with these magnetic fields \citation{NV3}.

It is also important to note that any auxiliary magnetic field used during the measurement process with NV centers is unlikely to significantly impact the dynamics of the topological insulator. This can be inferred by considering the interaction of the magnetic field with the edge electrons via the Zeeman effect, described by \(H_z = -g \mu_B \vec{\sigma} \cdot \bi{B}\), where \(\mu_B\) is the Bohr magneton and \(g\) is the g-factor of the edge electrons. The energy of the edge states is then expressed as \( E^{\eta} = \eta \sqrt{(\hbar v_F k + g \mu_B B_z)^2 + g \mu_B ( B_x^2 +  B_y^2)} \), with \(\eta = \pm\) corresponding to the conduction and valence bands. As long as the condition \(\hbar v_F \pi/a \gg eV \gg 2\mu_B |\bi{B}|\) is satisfied, the spin of the edge states and the resulting gap remain largely unaffected by the applied magnetic field. Notably, the auxiliary magnetic fields used in these measurements are typically two to three orders of magnitude smaller than the electric potential applied in this study \citation{NV3}, and thus minimizing any potential disturbances to the system's dynamics. 

\begin{figure}[t]
\centering
\noindent{\includegraphics[width=1\textwidth]{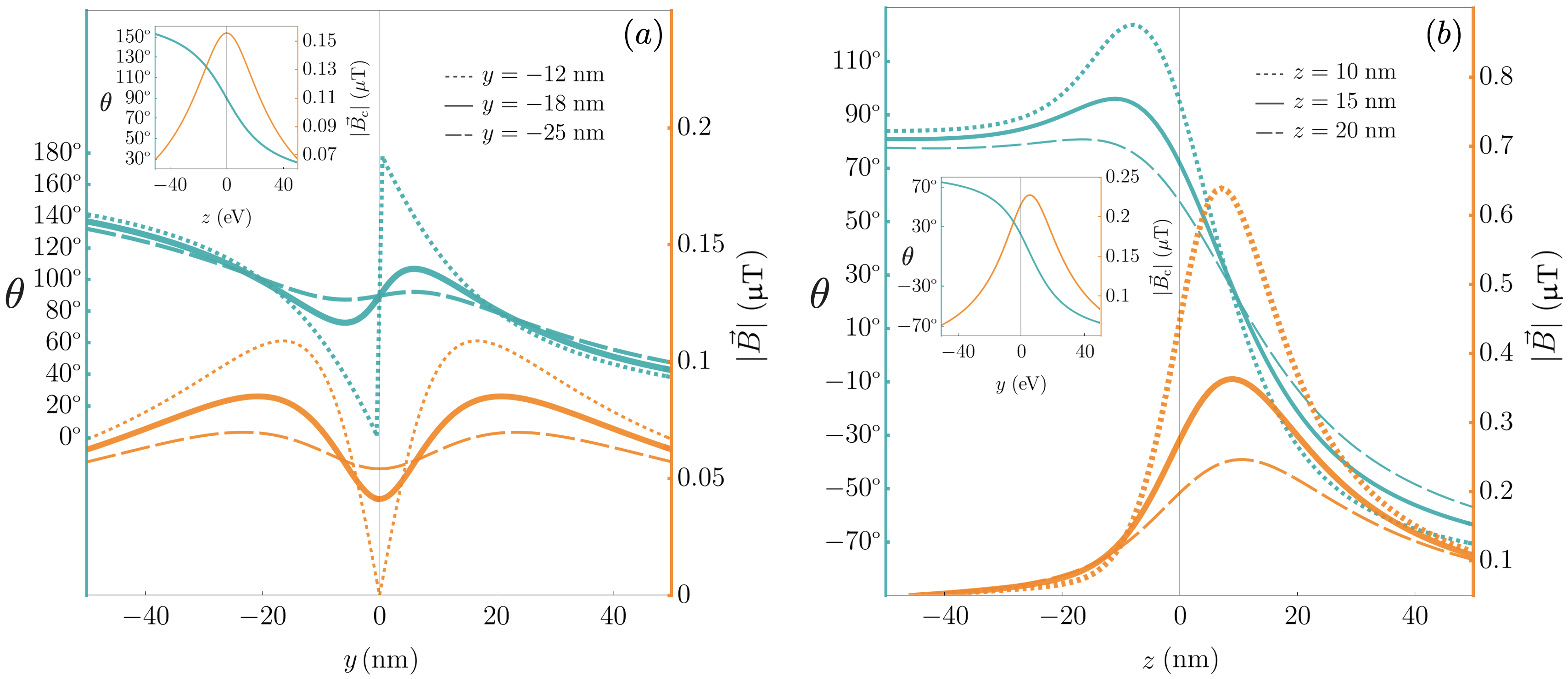}} 
\caption[]{Magnetic field magnitude (depicted in orange) and angle with respect to the $y$ axis (shown in cyan) as it traverses perpendicularly through the edge along the \(z\)-axis (a) and \(y\)-axis (b). The zero mark denotes the position of the edge. In both panels, an inset illustrates the magnetic field profile in the absence of the spin accumulation contribution. The fixed parameters for the topological insulator, consistent with the range of 2D topological insulators reported in the literature \citation{Reals_TI,Reals_TI1,TI_WTe}, are \(\textrm{Gap} = 0.1 \;\textrm{eV}\), \(\hbar v_F/a = 0.3 \;\textrm{eV}\), and \(C = -0.5 \;\textrm{eV}\:\textrm{nm}^{2}\), with a lattice constant of \(a = 3.5\:\)\r{A}.}\label{YZSwept}
\end{figure} 
\noindent

If the requirements for applying NV center magnetometry techniques are met, a complete profile of the edge magnetic field can be obtained \citation{NV3}. Figure \ref{YZSwept} illustrates the profile of the total magnetic field when sweeping along the \(y\)-axis with a fixed \(z\)-axis position and vice versa. Panel (a) shows the magnetic field profile along the \(z\)-axis for specific topological insulator parameters and an applied voltage of 1 mV. The point at which the magnetic fields cancel is clearly visible, and the inset reveals how this cancellation point vanishes if spin accumulation is not considered. In panel (b), the profile of the magnetic field is displayed for a sweep along the \(y\)-axis, with the inset showing the expected profile in the absence of spin accumulation. While the overall magnetic field magnitude shows minimal difference in this case, the field angle exhibits a notable peak just before entering the material from the left.

\section{Summary and Conclusions}\label{sec:five} 

In this work, we explore the magnetic field generated at the edge of a topological insulator in the QSH phase. Specifically, we examined a two-terminal device with two well-separated boundaries, where opposing magnetizations are induced at each edge by an electric current driven by a potential difference \( V \) that is two orders of magnitude smaller than the topological insulator’s bandgap. This setup reveals the helical edge states in the QSH phase, where opposite spin orientations at each edge produce a net spin accumulation, and thus an edge-specific magnetization detectable locally even though the overall system magnetization remains zero.

Using linear response theory, we calculated the magnetic field generated by the spin accumulation at the edge and assessed the distance at which this field becomes comparable to, or exceeds, that produced by the electron current. This distance  strongly depends on the topological insulator parameters, particularly the bandgap and the Fermi velocity. Our findings indicate that materials with larger bandgaps have more confined edge state near the boundary, making the angular component of the magnetic field vector measurable. Conversely, a larger Fermi velocity reduces the magnetic field magnitude associated with spin accumulation, which becomes detectable primarily near the edge. Notably, with a smaller gap (around 0.01 eV) or reduced Fermi velocities (\(\leq 1 \times 10^5 \) m/s), a favorable scenario arises where spin accumulation effects become detectable at distances exceeding 20 nm from the edge.

We propose that these fields can be measured using NV centers in nanodiamonds, ideally with diameters around 10 nm and coherence times of \(T_2 = 1~\upmu \textrm{s}\) as previously reported. Larger diameters could enhance sensitivity; however, the distance at which the magnetic fields due to spin accumulation and electron current are comparable generally shifts farther from the edge when the overall field magnitude is smaller. Consequently, our study concludes that it is feasible to detect the quantum spin Hall effect in topological insulators and to characterize the magnetic field near the edge using NV center magnetometry in nanodiamonds, with minimal impact on the intrinsic dynamics of the topological insulator.
 
\section{Acknowledgments}

The authors acknowledge the support from the Asian Office of Aerospace Research and Development (AOARD) FA2386-21-1-4125, Fondecyt Regular No 1221512, and ANID BECAS/DOCTORADO NACIONAL 21201120.

\appendix
\section{\label{sec:A} Magnetization calculation}

Consider the magnetization generated at the edge as described by the equation:
\begin{equation}\label{app_magnetization}
    m(\bi{r}) = \mu_{B}\left(\langle n^\uparrow(\bi{r})\rangle - \langle n^\downarrow(\bi{r})\rangle\right),
\end{equation}

\noindent where \( n^{\uparrow(\downarrow)}(\bi{r}) \) represents the charge density in a given direction, particularly at the upper edge (refer to Figure \ref{Current}). Due to the spin-locking of the edge states, each propagation direction is associated with a specific spin, reversed at the opposite edge. The bias voltage \( \phi(\bi{r}) \) alters the ratio between right and left movers due to this spin-locking mechanism. The total electron density can be expressed as:
\begin{equation}
n(\bi{r}) = \sum_{i=1}^{N} \delta(\bi{r} - \bi{r'_i}),
\end{equation}
\noindent which, in the basis of the Hamiltonian states for a specific edge, can be written in second quantization as:

\begin{equation}
n(\bi{r}) = \sum_{kk'\sigma\sigma'} \frac{1}{L} \int d\bi{r}' \, e^{-ik'x'} u_k'(y') \times \delta(\bi{r} - \bi{r'}) e^{ikx'} u_k(y') \langle\sigma'|\sigma\rangle c_{k'\sigma'}^{\dagger} c_{k\sigma}
\end{equation}

\begin{figure}[t]
\centering
\noindent{\includegraphics[width=.5\textwidth]{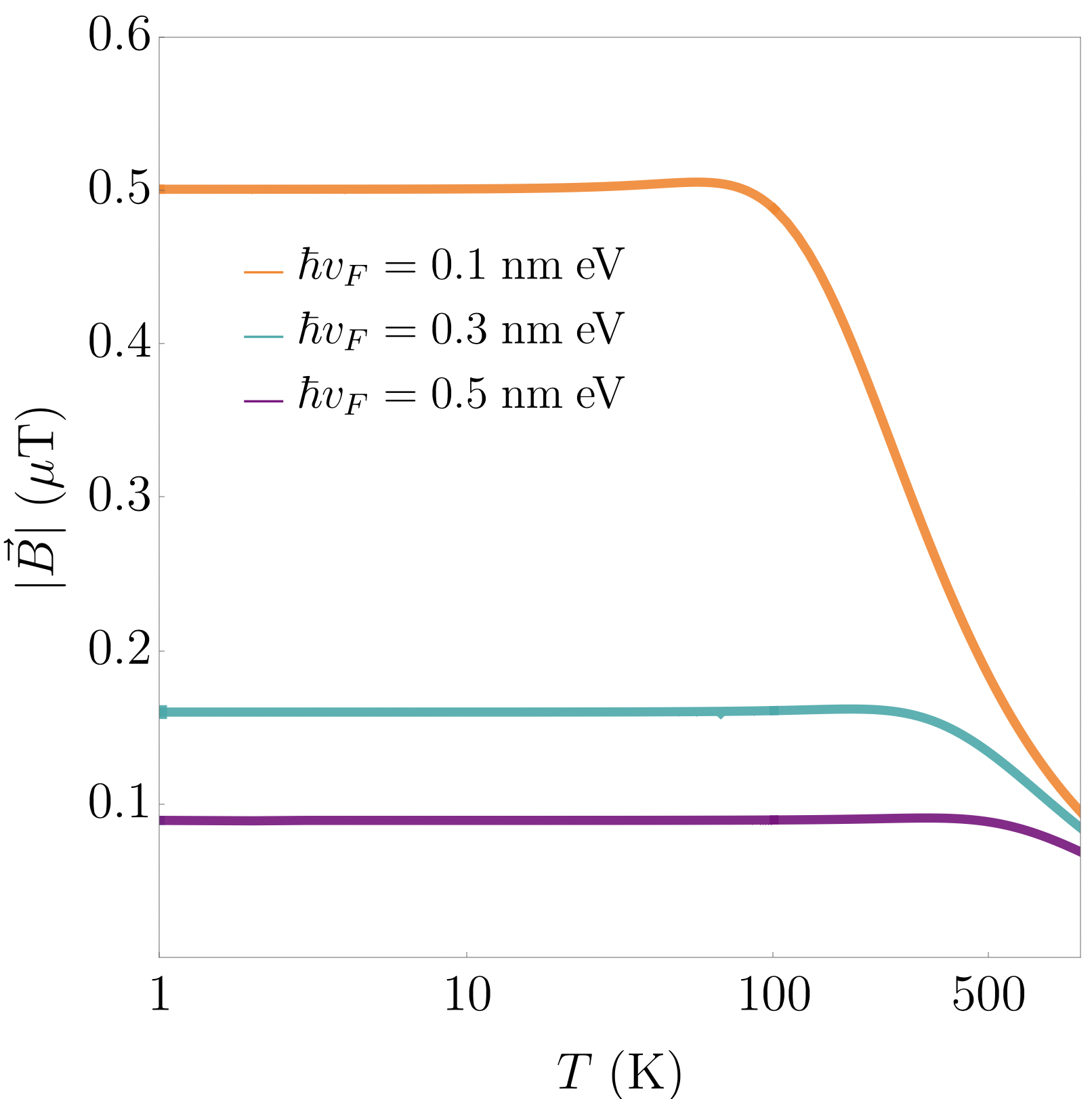}} 
\caption[]{Magnetic field due to spin accumulation as a function of temperature for different values of the Fermi velocity. The fixed parameters for the topological insulator are \(\textrm{Gap} = 0.05 \;\textrm{eV}\), \(\hbar v_F = 0.03 \;\textrm{eV}\:\textrm{nm}\), and \(C = -0.5 \;\textrm{eV}\:\textrm{nm}^{2}\), with a lattice constant of \(a = 3.5\:\)\r{A}. The measurement position is set at \(y = 15\:\textrm{nm}\), \(z = 15\:\textrm{nm}\).}\label{BvsT}
\end{figure} 
\noindent 

\noindent Focusing on the expectation value of the occupation number, we have:
\begin{equation}
\langle n^{\uparrow(\downarrow)}(y)\rangle = \sum_{k} \frac{1}{L} |u_{k'}(y)|^2 f^{\uparrow(\downarrow)}_{k},
\end{equation}
\noindent where \( f^{\uparrow(\downarrow)}_{k} = \left(e^{(E_{k}^{\uparrow(\downarrow)} \mp eV)\beta} + 1\right)^{-1} \) denotes the Fermi distribution, and we utilize the fact that:
\begin{equation}
\langle c_{k\uparrow(\downarrow)}^{\dagger} c_{k'\uparrow(\downarrow)}\rangle = \frac{\delta_{k,k'}}{\left(e^{(E_{k}^{\uparrow(\downarrow)} - \mu_{R(L)})\beta} + 1\right)},
\end{equation}

\noindent where \( \mu_{R(L)} \) is the chemical potential associated with each edge state on a given edge (see Figure \ref{Current}). Considering a weak electric potential, the first-order occupation number can be calculated as:

\begin{equation}
\langle n^{\uparrow(\downarrow)}(y)\rangle = \sum_{k} \frac{1}{L} |u_{k'}(y)|^2 \frac{df^{\uparrow(\downarrow)}_{k}}{d\mu_{R(L)}}\bigg|_{\mu_{R(L)}=0} \mu_{R(L)}.
\end{equation}
\noindent Thus, the magnetization expressed in equation (\ref{app_magnetization}) to first order is:
\begin{equation}
m(y) = \frac{\mu_{B}}{L} \sum_{k} |u_{k}(y)|^2 \left(\frac{df^{\uparrow}_{k}}{d\mu_{R}}\bigg|_{\mu_{R}=0} \mu_{R} - \frac{df^{\downarrow}_{k}}{d\mu_{L}}\bigg|_{\mu_{L}=0} \mu_{L}\right).
\end{equation}
\noindent Finally, considering the single-particle state energy \(E_{k}^{\uparrow(\downarrow)} = \pm v_F \hbar k\) and employing a continuum approximation, along with a semi-Fourier transform in the \(x\)-direction, the magnetization expression can be simplified as:
\begin{equation}
m(y,q) = \frac{\mu_B}{\pi}\frac{V e}{k_B T}\;\delta{(q)}\int\limits_{-\pi/a}^{\pi/a}dk\, |u_{k}(y)|^2  \frac{e^{v_{F}\hbar\: k\:/kB T}}{\left(1+e^{v_{F}\hbar\: k/kB T}\right)^2},
\end{equation}
which is the expression used in equation (\ref{main_magnetization}). It is important to note that the magnetization has been expressed in an approximate form to highlight its linear dependence on the electric potential. However, a more detailed analysis using the complete Fermi distribution shows that, for thermal energies lower than the gap between the conduction and valence bands of the topological insulator, no significant deviation is observed when compared to this approximation.

\newpage

\bibliographystyle{iopart-num} 

\end{document}